\documentstyle[epsf]{lamuphys}
\begin{document}
\title{In Situ Measurements of Interstellar Dust}
\author{M. Landgraf and E. Gr\"un}

\institute{Max-Planck-Institut f\"ur Kernphysik\\
D-69029 Heidelberg, Germany}

\maketitle
\begin{abstract}
We present the mass distribution of interstellar grains measured {\em in
situ} by the Galileo and Ulysses spaceprobes as cumulative flux. The
derived {\em in situ} mass distribution per logarithmic size interval
is compared to the distribution determined by fitting extinction
measurements. Large grains measured {\em in situ} contribute
significantly to the overall mass of dust in the local interstellar
cloud. The problem of a dust-to-gas mass ratio that contradicts cosmic
abundances is discussed.
\end{abstract}

\section{Introduction\label{intro}} 

Dust alters the conditions of the diffuse medium, e.g. it provides a
dominant heating mechanism via photo-electric heating~\cite{slavin97}.
So understanding the size distribution is important for modeling the
diffuse interstellar medium of which the local bubble is an example.
Extinction measurements give information about the size distribution
of grains in the diffuse medium and hints at their
composition~\cite{massa89}. The extinction curve can be fitted by a
mixture of graphite and silicate grains with a power-law size
distribution (MRN distribution) $n(a)\propto a^{-q}$ in the interval
$0.005\ {\rm \mu m}\leq a\leq 0.25\ {\rm \mu m}$ with an exponent of
$3.3\leq q\leq 3.6$~\cite{mathis77}. This optically active population
can explain all depletions~\cite{spitzer78}. No larger grains, that
are optically gray, are needed.  Interstellar dust was first detected
in situ by the dust detector on board the spaceprobe Ulysses in 1992
after Jupiter flyby~\cite{gruen93}. The detector, that is identical to
the one on board the Galileo spaceprobe, is an impact charge detector
and measures the mass, impact velocity, and impact direction of the
impacting grain as described by Gr\"un et al. (1992). The finding was
confirmed by the data collected by the Galileo dust detector. Galileo
reached Jupiter in December 1995 where planetary dust is dominant, so
the Galileo measurements range from mid 1993, when Galileo left the
inner solar system, to the end of 1995. Ulysses dust data is available
up to March 1996. The criteria to identify interstellar impacts in
both Ulysses and Galileo data were given by Baguhl et al. (1996). The
impact rate measured by Ulysses is $0.45\ {\rm day}^{-1}$ which
translates into a flux of $1\cdot 10^{-4}\ {\rm m}^{-2}\ {\rm
s}^{-1}$. Interstellar grains provide the dominant dust flux in the
outer solar system.

Analysis of the directional information~\cite{baguhl95} indicates a
stream of grains entering the solar system from a direction that is
compatible with the upstream direction of interstellar
helium~\cite{witte93}.

\section{Fit to Cumulative Mass Distribution\label{fluxfit}}
The grains detected {\em in situ} are larger than the ones needed to
explain the extinction curve. We check if the mass distribution is
just an extrapolation of the MRN population, ignoring the fact that
this would contradict cosmic abundance
considerations~\cite{spitzer78}. To avoid binning effects, we
investigate the cumulative mass distribution of flux $F(m)$. This is
the flux of grains with masses larger than a given mass.

\begin{figure}[ht]
\begin{center}
\epsfxsize=.6\hsize
\epsfbox{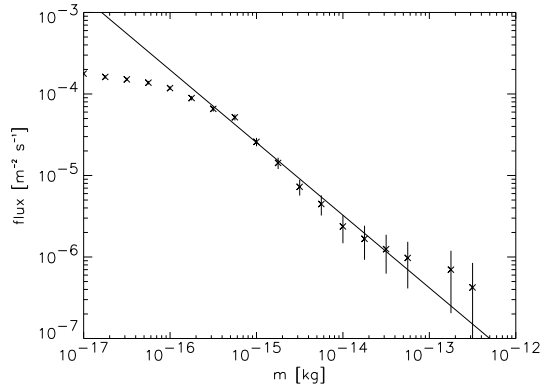}
\end{center}
\caption[]{\label{disfuncs}Cumulative mass distribution of grains
detected by Ulysses and Galileo including power-law fit. Error-bars
indicate statistical errors.}
\end{figure}

For the MRN distribution we expect $F_{\rm MRN}(m)\propto
m^{-q^\prime}$ with $q^\prime = 0.83$.  Fig.\ref{disfuncs} shows the
cumulative mass distribution of all particles detected by Ulysses and
Galileo together with the power-law fit to masses larger than
$10^{-16}\ {\rm kg}$, where the distribution was not altered by
filtration as described by Levy and Jokipii (1976). The fit is
weighted to match the points with good statistics.  An exponent of
$q^\prime = 0.90$ fits the data best with $\chi^2=9.4$. $\chi^2$ is
well below the $1\sigma$-limit of $\chi^2_{1\sigma,12} = 14$ for $12$
degrees of freedom. Due to bad statistics for grains with large
masses, the fit is not too sensitive, but indicates a steeper slope
than the MRN extrapolation. The $\chi^2$ for the MRN value is close to
the $1\sigma$-limit $\chi^2_{\rm MRN} = 14$. Slopes of $q^\prime > 1$
are ruled out on the $1\sigma$-level with
$\chi^2_{q^\prime>1}=20$. Such a steep dropoff would not lead to a
problem with cosmic abundances when extrapolating to large masses
since the small grains would then contribute more to the overall mass
than the large ones.

\begin{figure}[ht]
\begin{center}
\begin{tabular}{cc}
(a) & (b) \\
\epsfysize=5cm
\epsfbox{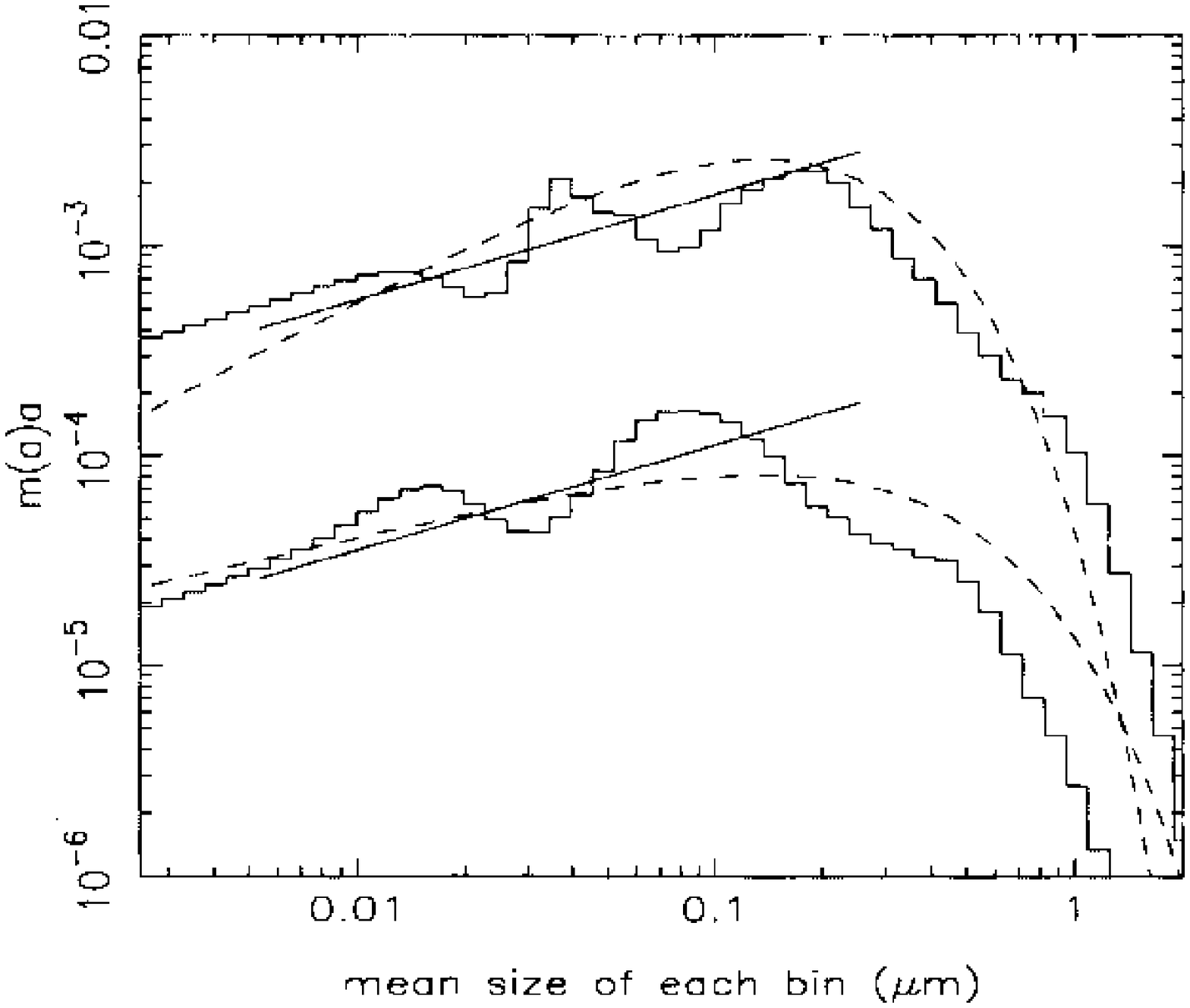} &
\epsfysize=5cm
\epsfbox{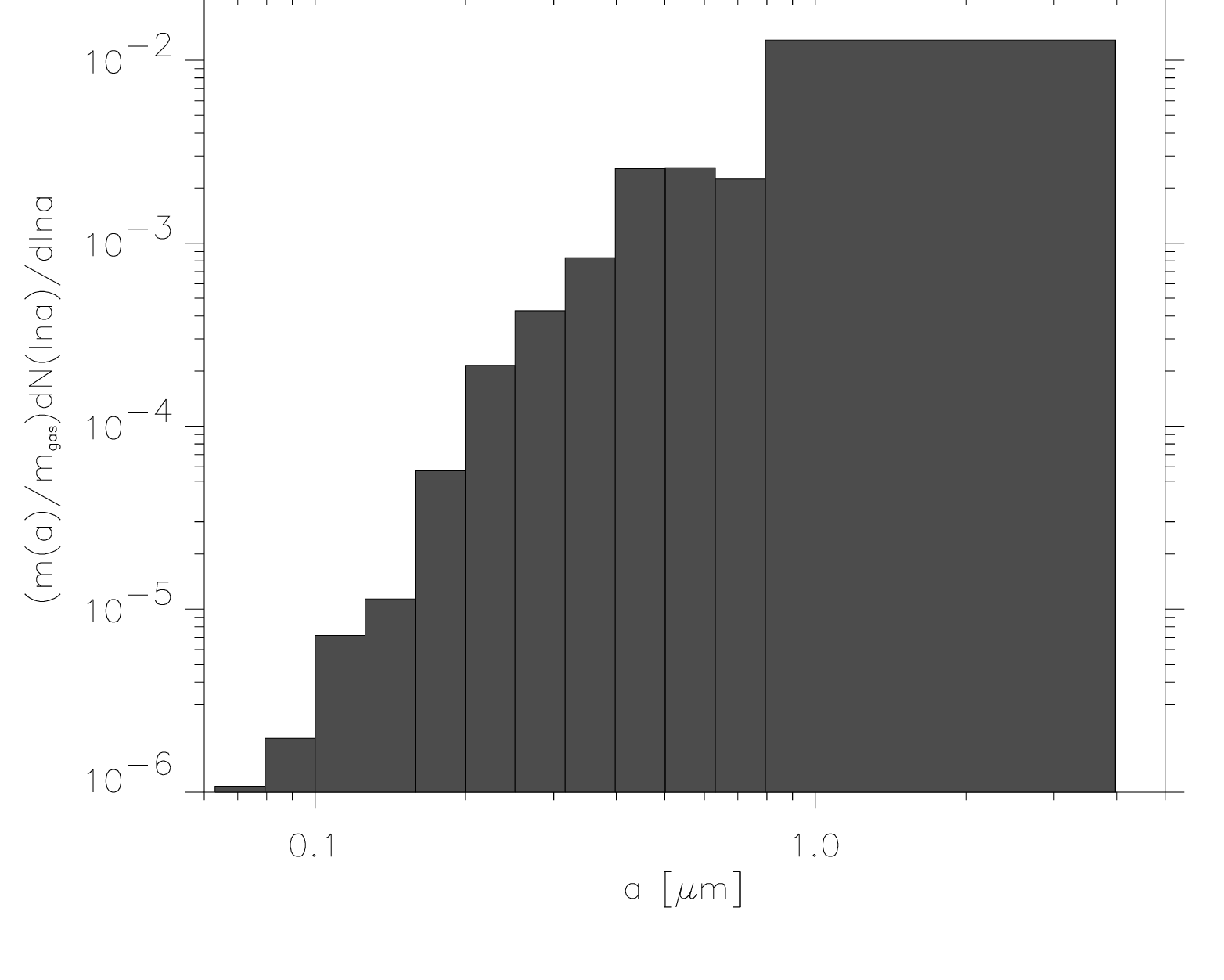}
\end{tabular}
\end{center}
\caption{\label{masssize} (a) Differential distribution of mass in silicate
(upper histogram) and graphite grains (lower histogram, scaled down by
a factor $10$) per logarithmic size interval normalized to the mass of
hydrogen in the same volume (taken from \protect\cite{kim94}, the
abscissa is the mean grain size in the bin and the ordinate is
identical to the ordinate in (b).). The
solid line is the MRN distribution and the dashed line is the PED-
(Power-law with Exponential Decay) fit. (b) Mass distribution per
logarithmic size interval of {\em in situ} data.}
\end{figure}

\section{Comparison to Extinction Data} 

We compare the {\em in situ} distribution to the distribution fitted
to extinction measurements by Kim et al. (1994). Since extinction is
measured along long lines of sight through the diffuse medium and {\em
in situ} measurements are very local compared to this, one has to be
careful interpreting this comparison. The fits to extinction
measurements rely on a dust model with a size distribution and
different grain compositions. Kim et al. (1994) give the mass per
logarithmic size interval normalized to the mass of hydrogen in the
same volume as shown in Fig.\ref{masssize} (a) assuming the canonical
value of $0.1\ \mbox{H-atoms}\ {\rm cm}^{-3}$ for the gas density in
the diffuse medium. We calculate the {\em size} distribution $N(\ln
a)$ from the {\em mass} distribution given in Sect.\ref{fluxfit} by
assuming homogeneous, spherical grains with a bulk density of
$\rho_{\rm ISP} = 2.5\ {\rm g}\ {\rm cm}^{-3}$~\cite{burns79}. From
the size distribution the differential mass distribution per
logarithmic size interval $(m(a)/m_{\rm gas}) \mbox{d}N(\ln
a)/\mbox{d}lna$, which is identical to the function given by Kim et
al. is calculated.
The comparison of Fig.\ref{masssize} (a) with Fig.\ref{masssize} (b)
shows, that the {\em in situ} distribution does not reproduce
the steep dropoff fitted to the extinction measurements. The {\em in situ}
distributions show, that there is much mass in grains larger than
$0.2\ {\rm \mu m}$.

\section{Discussion\label{discussion}} We have shown that the
existence of big ($a>0.2\ {\rm \mu m}$) interstellar grains is
evident. If the size distribution of small grains is extrapolated to
larger sizes as was indicated by Fig.\ref{masssize} (b), this will
lead to dust-to-gas mass ratios of more than the canonical value of
$1\%$, because $1\%$ is already used up by MRN grains and larger
grains contribute even more to the overall mass assuming the MRN
extrapolation. Dust-to-gas mass ratios larger than $1\%$ contradict
cosmic abundances~\cite{spitzer78}.
By integration of the distribution shown in Fig.\ref{masssize} (b),
the contribution of the in situ measured particles to the dust-to-gas
mass ratio is determined to be $2.2\%$. The problem of too much mass
in dust in the VLISM gets even worse if one extrapolates the size
distribution to even larger interstellar grains, which were identified
recently by Taylor et al. (1996) in radar meteor data. Of course the
extrapolation of a MRN-like power-law is limited otherwise the total
mass in dust gets infinite. 

Jones et al. (1996) show that large grains get destroyed in shocks
caused by supernovae. If the dust in the LIC has been shocked in the
past, than there should be much more smaller grains, and if not, the
dust has to be younger than the typical time-scale for returning the
dust mass into the gas phase by supernova shocks which is
$\tau_{\rm SNR}\approx 10^8\ {\rm a}$~\cite{jones96}.

\end{document}